\documentclass[aps,prd,preprintnumbers,nofootinbib,preprint]{revtex4}
\usepackage{bm}
\usepackage{latexsym}
\usepackage{dcolumn}
\usepackage{amsmath,amsfonts,amssymb}
\usepackage{graphicx,epsfig}
\usepackage{soul}
\usepackage{ulem}
\usepackage{amsthm}
\usepackage{color}
\begin{document}
%
%
%
\title{Renormalizing gravity: a new insight into an old problem}
%
%
%
\author{{Saurya Das}$^1$}\email[email:~]{saurya.das@uleth.ca}
\author{Mir~Faizal$^{^{2,1}}$} \email[email:~]{mir.faizal@uleth.ca}
\author{{Elias~C.~Vagenas}$^3$}\email[email:~]{elias.vagenas@ku.edu.kw}
%
%
\affiliation{$^1$ Theoretical Physics Group and Quantum Alberta, Department of Physics and Astronomy, University of Lethbridge, 4401 University Drive, Lethbridge, Alberta T1K 3M4, Canada}
\affiliation{$^2$ Irving K. Barber School of Arts and Sciences, University of British Columbia - Okanagan, 3333 University Way, Kelowna, British Columbia V1V 1V7, Canada}
\affiliation{$^3$Theoretical Physics Group, Department of Physics, Kuwait University, P.O. Box 5969, Safat 13060, Kuwait}
%

\begin{abstract}
\par\noindent
It is well-known that perturbative quantum gravity is non-renormalizable.
The metric or vierbein has generally been used as the variable to quantize in perturbative quantum gravity.
In this essay, we show that one can use the spin connection instead, in which case
it is possible to obtain a ghost-free renormalizable theory of quantum gravity.
Furthermore in this approach, gravitational analogs of particle physics phenomena can be studied.
In particular, we study the  gravitational Higgs mechanism using spin connection as a gauge field,
and show that this provides a mechanism for the effective reduction
in the dimensionality of spacetime.
\\
{}
\\
%
{}\\
{\bf 
{\it
This essay received an honorable mention in the 2018 Gravity Research Foundation Essay Competition on Gravitation
}}
\end{abstract}
%
%
\maketitle
%
General relativity (GR) and quantum field theory (QFT) are theoretically well-founded and
experimentally verified theories. Even though QFTs have divergences, they can be
dealt with 
by renormalization, for theories where the interactions are bounded by the free action \cite{hw}-\cite{hwhw}. 
In four dimensions, interactions are bounded by the free action in theories with quartic interactions terms because the first Sobolev norm bounds the volume integral of such interactions (as is the case of Yang-Mills theories) \cite{hw}-\cite{hwhw}.
Since all interactions in the Standard Model (SM) of particle physics are described by such theories, SM is renormalizable.
On the other hand, for perturbative quantum gravity (QG) in four dimensions,
governed by the action $S =  (1/{16\pi G_N})\int d^{4} x~\sqrt{-g}~{R}$
(where $R=$ curvature scalar, $G_N=$ Newton's constant),
the interactions are not bounded by the free action and thus perturbative QG is
non-renormalizable \cite{nonrenorm1}-\cite{sagnotti}. Furthermore, in the
above action, perturbation theory breaks down as the perturbations can
exceed the values of the original metric or vierbein \cite{hw}-\cite{hwhw}.
The situation changes, however, if one adds higher curvature terms to the action, namely
\begin{equation}
 S =  \int d^{4} x~\sqrt{-g}
 \left[-\frac{R}{16\pi G_N}  - \alpha R_{\mu\nu}R^{\mu\nu} + \beta R^2\right]~.
\label{act1}
\end{equation}
In this case, although one has interactions greater than quartic, the theory is still
renormalizable \cite{6}, as now
the second Sobolev norm bounds the pointwise value of perturbations, and
hence the free action bounds the interactions \cite{hw}-\cite{hwhw}.
%
%
However, it is well-known that this theory contains Ostrogradsky ghosts, giving rise to
negative norm states and negative probabilities \cite{aq}-\cite{bq}.
The origin of these ghosts is the presence of higher derivative terms, which occur
in this theory when the curvature scalar $R$ is expressed in terms of the metric ($g_{\mu\nu}$) or the vierbein
($e^{a}_{\mu}$)\footnote{Greek letters denote spacetime indices and run from $0$ to $d$, while Latin letters are used as tangent space indices and run from $1$ to $d$, in a $(d+1)$-dimensional spacetime.}.

In this essay, we show that the above problem does not occur if one uses the spin connection ($\omega^{ab}_{\mu}$)
as the variable to quantize. It may be noted that {the}
spin connection has also been used in Loop QG \cite{loop}.  However they have not been
studied in detail in
perturbative QG. Note that in classical gravity the use of the spin connection gives rise to identical predictions as with
metric variables, including experimentally measurable (gauge invariant) quantities. However their behavior in quantum theory can be quite different.

In this picture, gravity can be considered as a gauge theory with the
spin connection as a $SO(3,1)$ gauge field \cite{ramond}.
The metric is given by $g_{\mu\nu} = e_\mu^a e_{a\nu}$ with $e=\sqrt{|g|}$.
Therefore, the curvature tensor can be written as the field strength of the spin connection
%
$ R_{\mu\nu}^{ab}  \equiv F_{\mu\nu}^{ab} =
\partial_{[\mu} \omega_{\nu]}^{ab} - \omega_{[\mu}^{ca} \omega_{\nu] c}^b~.
$
%
%
Then the curvature scalar
is $R=R(\omega)$, while the action is $S=S[e,\omega]$ (the $e$ comes from the measure).
The expression for the spin connection in terms
of vierbeins
$\omega_\mu^{ab}= -e^{\nu a}\nabla_\mu e^b_\nu$
and the Einstein equations are obtained by
varying $S$ with respect to $\omega_\mu^{ab}$ and
$e_\mu^a$ respectively \cite{ramond}.
Next, one can write a higher curvature action for gravity as a topological field theory coupled to Yang-Mills theory in four dimensions as follows
\begin{eqnarray}
 S &=&
 \int d^{4} x~e
 \left[ - \frac{B^{ab}_{\mu\nu}F^{\mu\nu}_{ab}}{16\pi G_N}  - \frac{1}{4g^2} \, F^{\mu\nu}_{ab}
 F_{\mu\nu}^{ab}   \right]
 \label{act3}
\end{eqnarray}
with $B_{\mu\nu}^{ab} = e_{\mu}^a e_{\nu}^b $
and $g$ is a coupling constant. It may be noted that in this formalism,
Einstein gravity (the first term) is a topological theory
similar to a BF theory \cite{bf}.
Furthermore, the second term is of the form of the higher curvature terms in
Eq.(\ref{act1}), using the fact that
$ \int d^4 x~e \left[ R_{\mu\nu}^{ab} R^{\mu\nu}_{ab}-
4 R_{\mu}^a R^{\mu}_a +R^2\right]
$
is the (Gauss-Bonnet) topological invariant in four-dimensions, and vanishes for a topologically trivial background.
It may be noted that even though consistency of Newtonian limit of this theory with observational data needs further investigation, this theory is clearly a ghost-free renormalizable theory in these variables.
This is because, as seen from Eq.(\ref{act3}), the interactions therein are quartic, just as in the case of SM.
%
%
%
%
%
%
Here instead of perturbing the metric or vierbein, one now studies the theory by perturbing the spin connection
\begin{equation}
 {\omega}_{\mu}^{ab} =  \tilde{\omega}_{\mu}^{ab} + \bar{\omega}_{\mu}^{ab}~.
 \end{equation}
where $\bar{\omega}_{\mu}^{ab}$ is the background, and one
quantizes the fluctuations $\tilde{\omega}^{ab}_\mu$ around it.
This can then be used to compute scattering processes involving gravity, now described in terms of these fluctuations.

Having {proposed}
a ghost-free renormalizable theory of gravity as a gauge theory,
one can explore gravity analogs of SM phenomena.
In particular, in what follows, we study one such example, namely spontaneous symmetry breaking (SSB) or the Higgs mechanism in gravity.
Gravitational Higgs mechanism has been studied in the past in various contexts,
including in a few instances using the spin connection
\cite{earlier}-\cite{ealier1}, but none with the aim of spontaneous dimensional
reduction, to the best of our knowledge.
%
%
%
For generality, we will analyze SSB
in $(d+1)$-spacetime dimensions.
We start with a Higgs field $\Phi$
which transforms under the {\it vector} representation of the
$(d+1)$-dimensional Lorentz group
(note that the vector representation
was also used in the Georgi-Glashow model for weak interactions \cite{georgi}).
Then, $\Phi$ has $(d+1)$ real components.
We choose a non-negative Higgs potential of the form
\begin{equation}
V(\Phi^\dagger\Phi) = \frac{m^2}{\phi_0^2}
\left[ \Phi^\dagger\Phi - \phi_0^2 \right]^2 \geq 0~,
\label{vphi}
\end{equation}
where $m$ and $\phi_0$ are constants.
The vacuum minimizing the above potential is $\Phi=(0,\dots,\phi_0)^T$.
Fluctuations around this vacuum are denoted by $h$, i.e.
$\Phi=(0,\dots,\phi_0+h)^T$.
Next, using the
covariant derivative  $D_\mu\Phi= (\partial_\mu + \frac{i}{2}\omega_\mu) \Phi$,\footnote{Here,
$\omega_\mu =\omega^{ab}_\mu~\Sigma_{ab}$ with
 $\Sigma_{ab} = -i[\gamma_a, \gamma_b]/4$
and
$ [D_\mu, D_\nu] = 
\frac{i}{2}R^{ab}_{\mu\nu} \Sigma_{ab}$ .}
the Lagrangian for the Higgs field coupled to gravity via spin connection
can be written as follows, with subsequent expansions around the vacuum
\begin{eqnarray}
\mathcal{L}
&& = (D_\mu \Phi)^\dagger (D^\mu  \Phi)
- V(\Phi^\dagger \Phi) \nonumber \\
&& =
\frac{\phi_{0}^2}{4}
\left[
\omega_\mu^{0d}~\omega^{0d \mu}
+ \omega_\mu^{1d}~\omega^{1d \mu}
+ \dots
+ \omega_\mu^{d-1,d}~\omega^{d-1,d \mu}
\right] \nonumber \\
&& + 2m^2 h^2
+ \frac{1}{2}~\partial_\mu h \partial^\mu h +
\left[ \frac{h^2}{8} + \frac{h\phi_{0}}{2\sqrt{2}}
\right] \times
\label{scalar4} \\
&& \left[
\omega_\mu^{0d}~\omega^{0d \mu}
+ \omega_\mu^{1d}~\omega^{1d \mu}
+ \dots
+ \omega_\mu^{d-1,d}~\omega^{d-1,d \mu}
\right]  \nonumber \\
&& + \frac{m^2}{\phi_0^2}
\left[ \sqrt{2}~\phi_0~h^3 + \frac{1}{4} h^4 \right]~.\nonumber
\end{eqnarray}
%
%
%
%
It is seen from the above,
that the $d$ spin-connections, namely $\omega_{\mu}^{ad}$, with $a=0,\dots,d-1$
have each acquired a mass $M_\omega= \phi_0/\sqrt{2}$. The corresponding interactions are
therefore short-ranged.
The Higgs field also acquires a mass $m$.
The remaining $d(d-1)/2$ spin connections
remain massless, accounting from the long-ranged
nature of gravity.
The symmetry of the theory
spontaneously reduces from:
$SO(d,1) \rightarrow SO(d-1,1)$.
One can easily show that
the total number of
degrees of freedom (d.o.f.)
before and after SSB is the same.
Before SSB, one has
$(d-1)$ d.o.f. for each of the $d(d+1)/2$ massless spin connection and one for each of the $(d+1)$ scalar components,
i.e. a total of
$d(d+1)(d-1)/2+(d+1)=(d^3+d)/2 + 1$ d.o.f.
After SSB, one adds up the
d.o.f.
for the $d$ massive spin connections
(each with $d$ d.o.f.),
$d(d-1)/2$
massless spin connections
(each with
$(d-1)$ d.o.f.)
and one residual massless scalar field. This results in
$d\times d + d(d-1)/2\times (d-1) +1 =(d^3+d)/2 + 1$
exactly same as before.
%

It is natural to equate the SSB scale to the
Planck scale in $d$-spacetime dimensions,
such that the mass acquired by massive spin connections is of the order of Planck mass $M_{pl}^{(d)}$.
This implies that they cannot be accessed by low-energy phenomena,
and the dynamics in spacetime is effectively described by a lower, $d$-dimensional theory.
In other words, SSB has caused an effective dimensional reduction from $(d+1) \to d $ dimensions.
%
This mechanism provides an alternative to Kaluza-Klein
compactification as a means of dimensional reduction,
and may have potential applications in string/M-theory for dimensional reduction from ten/eleven-dimensions to the observed four-dimensions.
%
%

%

We conclude from the above that
gravitational Higgs mechanism presents a viable
method for the
emergence of the observed $4$-dimensional spacetime from a $5$-dimensional one,
near the Planck energy scale $M_{Pl}c^2 \approx 10^{16}~TeV$. 
Our earlier comments about the renormalizability of gauge and
gravity theories continue to hold in the final four-dimensional theory.

It is worth noting that in this case of $5\rightarrow 4$ dimensional reduction via SSB,
for matter coupled to the residual {\it massive} spin connections,
the dimensional Newton's constant $G_N$ appears naturally from a dimensionless coupling $\lambda$.
The easiest way to see this is to start with a matter
current $j_{ab}^\mu$ coupled to the massive spin connections via an interaction Lagrangian,
%
%
%
$
\mathcal{L}_{int}= - \lambda
\sum
j_{ab}^\mu
~\omega_\mu^{ab},
\label{lambda1}
$
where the sum 
is over the massive spin-connections.
Then, at energies below $M_\omega=M_{Pl}$,
the mass terms in Eq.(\ref{scalar4})
dominate and
the effective Lagrangian is given by
%
%
$
{\mathcal L}_{ eff}=
\lambda \sum 
\left[\frac{1}{2} M_\omega^2~\omega_{ab\mu}\omega^{ab\mu}
- \lambda j_{ab \mu} \omega_{ab}^{\mu} \right].
$
%
Varying 
this with respect to $\omega_\mu^{ab}$, and substituting the stationary solution
$\omega_{ab \mu} = \frac{\lambda}{M_\omega^2} j_{ab \mu}~$
back in the effective action yields
%
\begin{eqnarray}
{\mathcal L}_{ eff}
&=&  -\sum
G_{N}
~j^{ab}_{ \mu} j_{ab}^\mu\hspace{2ex}\mbox{with}\hspace{2ex}
G_{N}
\equiv \left(\frac{\lambda}{\sqrt{2}M_\omega} \right)^2~.
\end{eqnarray}
%
%
This phenomenon is similar to the emergence of
the effective dimensional Fermi constant $G_F$ from the
dimensionless $SU(2)$ coupling constant $g_2$, namely
$G_{F}= g^{2}_{2}/4\sqrt{2}M^{2}_{W}$ with $M_W$ as the $W$ boson mass
\cite{cottingham}-\cite{chengli}.
Note that this only occurs for massive spin connections, and not for the massless long-ranged spin connections.

In summary, we have shown here that 
one can study perturbative QG using spin connections as the dynamical variable, and that
a higher curvature theory of gravity written in terms of these connections
gives a ghost-free and renormalizable theory of QG.
Furthermore, it is possible to study gravitational analogs of phenomena in particle physics in this picture.
In particular, we have studied SSB due to gravitational Higgs mechanism.
%
Further work in this direction may include computing corrections to Newton's law and other scattering processes involving particles interacting via spin connections to show that they are finite.

%
%

\vspace{0.2cm}
\noindent
{\bf Acknowledgment:}
This work was supported in part by the
Natural Sciences and Engineering Research Council
of Canada and the University of Lethbridge.
%
%
%

%
%
%
%
%
%
%
%
%
%
\end{document}